# Trust, but verify: Reducing artificial light emissions and monitoring compliance

Salvador Bará

*Former profesor titular (retired) at Universidade de Santiago de Compostela (USC), Santiago de Compostela, 15782 Galicia (Spain, European Union)*

e-mail: salva.bara@usc.gal

This preprint is an extended version of the talk with the same title presented at the 3rd International Conference on Environmental and Astronomical Light Pollution EALPO 2024. Niepołomice, Poland. September 6th-8th 2024.

---

**Abstract:** In order to reduce light pollution we have to reduce the overall amount of artificial light emissions. This is a consequence of the basic rules governing the propagation of light in the terrestrial environment. In this work I revisit the physical laws causing that (i) "all" artificial light emitted outdoors is lost or pollutant, (ii) the negative effects of light pollution depend monotonically on the local concentration of photons of anthropogenic origin, and (iii) this concentration depends linearly on the total artificial light emissions, weighted by the light pollution propagation functions. Setting total emission limits becomes necessary in order to ensure that the negative effects of light pollution do not surpass red-lines of unacceptable degradation of the natural night. Once these red-lines are socially agreed, monitoring compliance becomes a relevant task. Several complementary methods for assessing total light emissions are being used nowadays, including public inventories of installed lights, direct radiance measurements from ground or low Earth orbit satellites, and scattered radiance measurements using ground based detectors (night sky brightness monitoring). While updated administrative inventories could in principle be trusted, independent verification is a must. The required measurements pose in turn significant challenges: we discuss here how the variability of the terrestrial atmosphere sets a lower limit on the minimum emission changes that can be reliably detected by measurements, and propose some ways to improve this performance.

---

## 1. Introduction

Light pollution is an interdisciplinary field of research whose outcomes are relevant for the formulation of public policies. Artificial light, no doubt one of the most useful technological inventions of humankind, has also proven to be a powerful environmental pollutant. Its presence at nighttime in places that would otherwise be lit only by natural sources is reported to cause complex disturbances at multiple scales of life (Longcore and Rich 2004; Rich and Longcore 2006; Hölker et al. 2010; Gaston et al. 2013). Light carries energy and information, triggering important physiological processes and providing essential environmental cues through different visual and non visual pathways. The natural patterns of light and darkness that shaped the evolution of life in the surface of our planet are presently under threat in wide regions the world (Falchi et al. 2016; Garret et al. 2020; Cox and Gaston 2023), including the photic zone of oceans and freshwaters (Perkin et al. 2011; Davis et al. 2013, 2014, 2016; Smyth et al. 2021, 2022; Tidau et al. 2021; Pérez-Vega et al. 2024).

Artificial light at night fulfills all conditions to be considered a classical pollutant (Bará and Falchi 2023). Light is composed of particles (Einstein 1905), and light pollution can be conceptualized as an increase of the concentration of light particles in the environment, due to human activity, that modifies the expected natural illumination levels and produces, or is able to produce, detrimental consequences on life, human health, and other tangible and intangible goods (United Nations 1996, 2018; Cinzano and Falchi 2012; Bará et al 2022) .

From a historical perspective, the use of artificial light has increased at a remarkable rate, especially since the generalized introduction of electric lighting in industrialized countries in the first decades of the 20th century (Fouquet and Pearson 2006, 2012; Tsao and Waide 2010). The widespread deployment of electric outdoor lighting in cities and roadways connecting them was strongly tied to the development of urban mobility patterns based on the massive use of private vehicles running at high speed throughout the urban fabric. This resulted in an escalation of the average illumination levels, not always based on verifiable science (Fotios and Gibbons 2018). This industry-driven trend is expected to continue. Since the human visual system may adapt to work under a very wide range of lighting levels, from a tiny fraction of lx to well above $10^5$ lx, the amount of artificial light present in our world at night might still potentially grow by several orders of magnitude, only limited by the fluctuating prices of energy. This increase is facilitated by the shifting baseline syndrome (Pauly 1995; Lyytimäki 2013), related to environmental generational amnesia (Kahn 2002; Kahn and Weiss, 2017; Bogard 2013), whereby each new generation considers "normal" the progressively higher levels of light they get acquainted with when they first "discover the world", leading to a very fast degradation of the night in the course of a few generations.

Contrary to some widespread belief, the increase of light pollution cannot be stopped by just improving the lighting installations. Light pollution is produced not only by bad installations but also, and to a great extent, by ideal, perfect ones. Light itself is the pollutant, independently of where it comes from. While this problem is naturally aggravated in case of badly designed or implemented lighting projects, it necessarily persists in technically perfect ones: the reason is that practically all light reflected in the surfaces we need to illuminate is lost or sent towards the natural environment, because it is not captured by any viewer (section 2). This is a fundamental constraint of nature that cannot be avoided by techno-optimist approaches aimed to improve the performance of the individual luminaires and the overall efficiency of outdoor lighting systems. In order to keep at bay the negative effects of artificial light at night in the environment it is ultimately necessary to put a limit on the total light emissions of the surrounding territory (Bará et al. 2021). Since in many places of the world the admissible deterioration of the night has already been surpassed, a planned reduction of the present emissions becomes the advisable way forward (Falchi et al. 2016, 2023; Falchi and Bará 2023).

The structure of this paper is as follows. In section 2 the physical principles supporting the above statements are briefly revisited. By how much should we effectively reduce the overall light emissions is discussed in section 3. The challenges posed by the need of verifying the actual outcomes of public lighting policies deserve their own space, and are described in section 4. Additional remarks are drawn in section 5.

## 2. The physical principles of light pollution control

Any effective strategy for light pollution control shall take into account the basic physical principles governing the propagation of artificial light. Although science by itself does not determine the approach that should be chosen to solve a sociotechnical problem (this is the realm of ethics and politics), the adopted solutions shall always be consistent with science, not contradictory with it. From a physical standpoint, there are several basic light pollution facts that should be kept in mind:

### *2.1. Practically all artificial light emitted outdoors is lost or pollutant*

it is frequently overlooked that practically all light we emit outdoors is lost or pollutant. Crucially, this is not due to technical defects in the lighting installations. It also applies to perfect, ideal ones. It is a direct consequence of the way light particles interact with most of the surfaces found in nature and in anthropized spaces.

The light we use to illuminate urban structures (pavements, sidewalks, façades...) is partly absorbed by their surfaces and partly reflected diffusely from them. Absorption is always present and represents a net loss of energy, although it can be reduced to some extent by tailoring the reflectance of the urban materials. Diffuse reflections, in turn, have a triple soul: they are at the same time useful, unavoidable, and the main cause of the intrinsically small efficiency of outdoor lighting systems.

Diffuse reflections are useful for vision because the reflected light propagates in all directions, allowing any illuminated point to be seen by observers located anywhere around it. They are also unavoidable, due to the roughness of the surfaces of the materials available in our natural and built environment (only mirror-like, polished surfaces reflect in well-defined directions the light from a distant point source). And they are the reason why only an exceedingly tiny fraction of the reflected light is actually used for vision. The human eye can only "see" the photons that propagate in the direction of its pupil and make their way through the ocular media to be absorbed by the retinal photoreceptors. The small size of the human pupil severely limits the fraction of reflected light that can effectively enter the eye.

For isotropic reflections (equal radiant intensity reflected in all directions, from each elementary patch of the surfaces), the fraction of reflected photons captured by an eye is equal to the ratio of the solid angle subtended by the pupil as seen from the reflection point to the solid angle of all possible reflection directions ($2\pi$ steradian, sr). The solid angle subtended by the pupil, in turn, is equal to the projected pupil area divided by the square of the distance to the reflection point. A conservative estimate for generously dilated pupils and illuminated points located at typical urban distances from the observer (~10 m) shows that only 1 photon out of every 22+ million reflected at such distance is captured by the eye (Bará and Falchi, 2023). A formal calculation extended to all points surrounding the observer, from their feet to several hundred meter away, taking into account the dependence of the pupil solid angle on the squared inverse distance and on the cosine of the angle of incidence of the light rays, results in even smaller overall fractions. If the diffuse reflections are not isotropic but Lambertian, further reductions of the fraction of captured photons are to be expected, due to the dependence of the reflected Lambertian radiant intensity on the cosine of the normal angle, a cosine which tends to be small for human observers looking at urban surfaces from typical viewing distances.

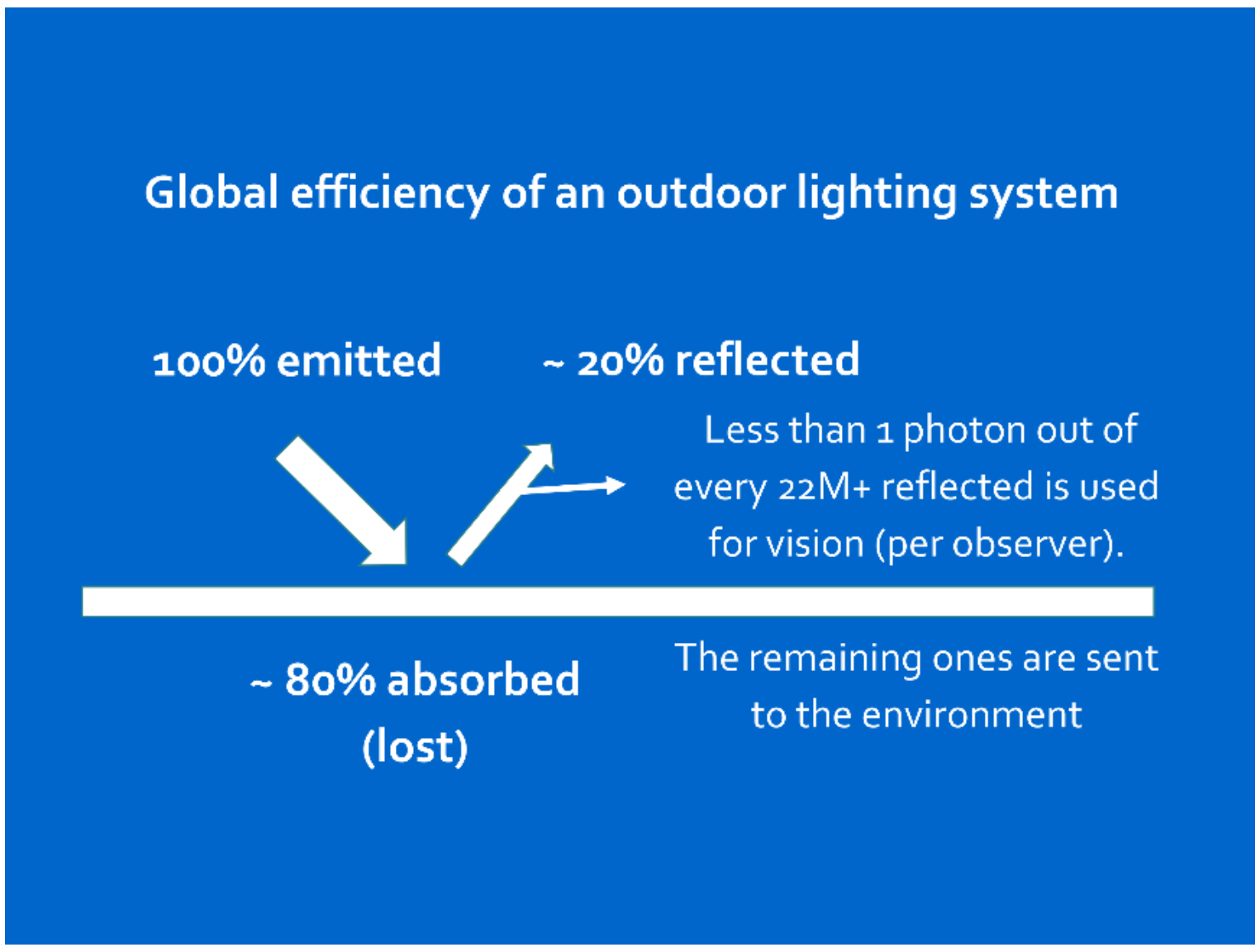

**Fig. 1.** The maximum overall efficiency of an outdoor lighting system illuminating common material surfaces of our urban and natural environment.

This "1/22M+" ratio means that the maximum efficiency attainable by an ideal, perfect artificial lighting system, measured as [total number of captured photons]/[total number of emitted photons], is smaller than 0.000005 % per observer present in an illuminated outdoor space with 100% surface reflectance. In other words, the 99.999995 % of the light reflected in their surroundings cannot be seen by that person. For usual urban surfaces with average reflectance of order 20%, the above overall efficiency (captured/emitted photons) becomes five times smaller: 0.000001 % per observer, so that 99.999999 % of the emitted light is not used for vision by that person (Fig.1). The overall efficiency increases in proportion to the number of observers looking to the same point on the street at the same average distance, but the resulting value is still very small for any imaginable number of actual observers. The huge fraction of reflected light not captured by any eye –practically all, in environmental terms– propagates away from these surfaces, producing light pollution effects at sites located at near and large distances.

The "1/22M+" ratio is a physical limit that cannot be overcome by technological means in practical outdoor lighting systems. It is not due to any flaw in the installations. It arises due to the way light interacts with the surfaces we illuminate to be seen. It applies also to ideal installations: once the light is successfully put where and when required (utilization factor equal to unity, presence detectors, etc.), with minimum illumination levels and appropriate spectra, the diffuse reflections unavoidably take their lead. Incorrectly designed or badly implemented lighting installations just make the situation worse.

### *2.2. The unwanted effects of light pollution depend monotonically on the local levels of light*

Electromagnetic radiation impacts living beings through several interaction mechanisms. The main interactions in the visible region of the spectrum are photochemical. The absorption of light particles initiates or facilitates chemical reactions that elicit a wide range of physiological and behavioral responses through different visual and non-visual pathways. Photosynthesis, light-induced oxidative processes, and vision are examples of this type of interaction.

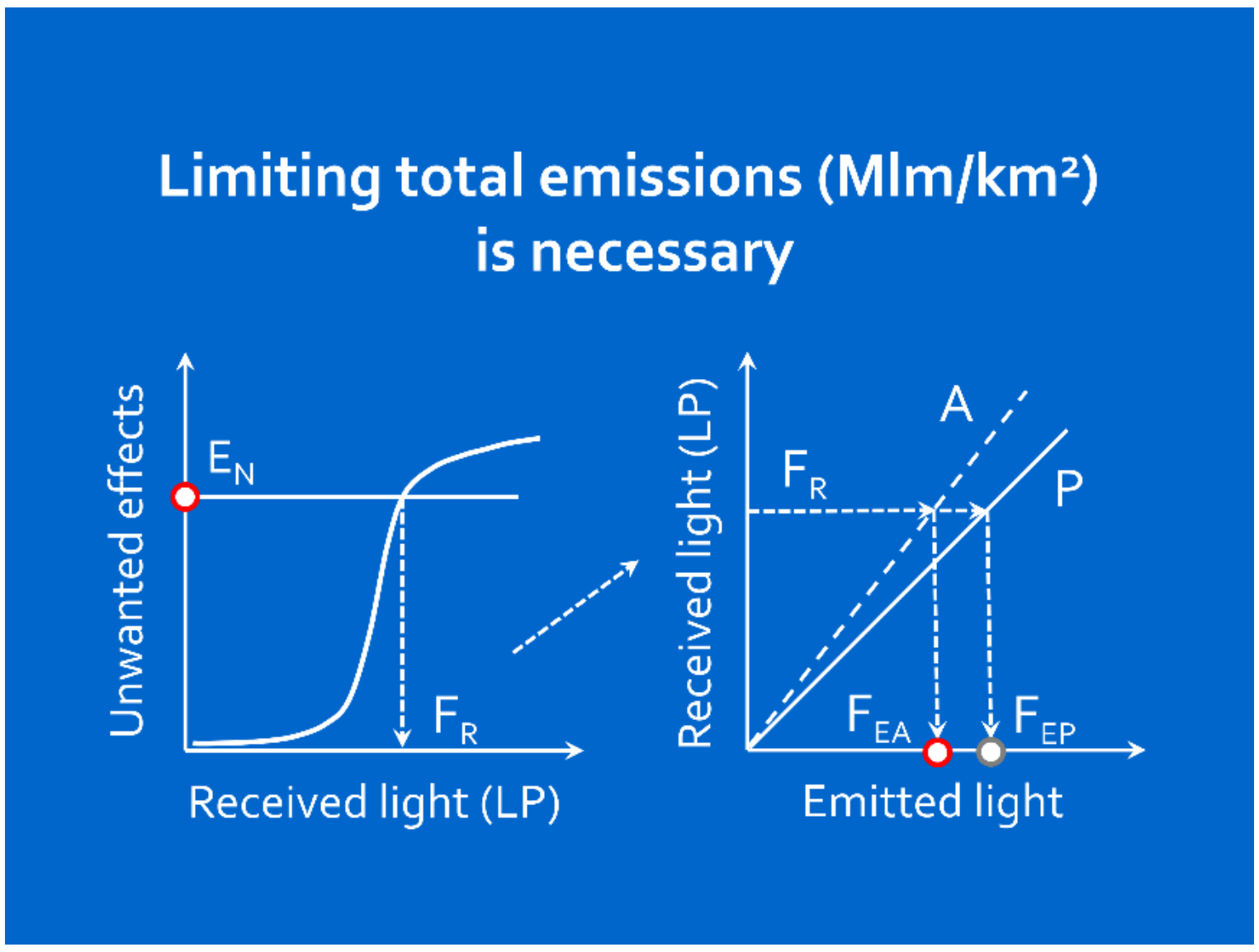

**Fig. 2.** *Left*: The strength of the unwanted effects of artificial light at night depends in a non-linear, monotonic way on the local levels of light. *Right*: The local levels of light, in turn, depend in a linear way on the total emissions of light in the surrounding territory, up to distances that can reach several hundred km. In order to ensure that the unwanted effects do not surpass a given red-line $E_N$, the total emissions of light shall not surpass a specific amount ($F_{EA}$ for practical lighting systems, $F_{EP}$ for ideal, perfect ones).

A typical feature of these interactions is that the strength of their effects depends in a non-linear way on the exposure to light. The functions describing this behavior often have an approximate sigmoid shape, as shown in Fig. 2 (left panel). Irrespectively of the details of each particular interaction, the crucial point is that this dependence tends to be monotonically increasing, i.e. larger exposures produce larger effects. For low levels of light exposure the effects increase at a slow pace, grow faster for intermediate levels, and saturate for high levels. Examples of this dependence can be found in the classical works of Brainard et al. (2021) and Thapan et al. (2021) on light-induced melatonin suppression.

This monotonic behavior has a relevant corollary with practical consequences for public policy-making: in order to ensure that the unwanted, negative effects of the exposure to artificial light at night do not surpass a given red-line value, e.g. $E_N$ in Fig. 2 (left), it is necessary to ensure that the local levels of light, in turn, do not surpass a maximum value $F_R$ which depends on $E_N$. As shown in next section (2.3), the local levels of light depend linearly on the total light emissions (Fig. 2, right). Taken together, the above statements mean that in order to not surpass the $E_N$ red-line, the total emissions shall be kept below a maximum value $F_{EA}$, for actual lighting systems (dashed line A). Perfect, ideal outdoor lighting systems (full line P) could encompass a slightly larger maximum amount of emissions $F_{EP}$ without crossing the red-line. However, once a system reaches its ideal status, emissions cannot be allowed to grow any further.

### *2.3. The local levels of light depend linearly on the total light emissions in the surrounding territory*

The propagation of light, at the usual irradiance levels of artificial outdoor lighting sources, is a linear process. This means that the illumination produced by a set of light sources is equal to the sum of the illuminations that each source would produce separately. In other words, the total amount of artificial light reaching the observer is a weighted sum of the sources' emissions. This can be abbreviately expressed as

$$B(\mathbf{r}) = \int K(\mathbf{r}, \mathbf{r}')\, L(\mathbf{r}')\, \mathrm{d}^2\mathbf{r}' \qquad (1)$$

where $B(\mathbf{r})$ is any linear light pollution indicator evaluated at the location $\mathbf{r}$, for example the horizontal illuminance or the artificial sky radiance (Duriscoe 2016; Falchi and Bará 2021; Bará et al 2022b; Falchi et al 2023b), $L(\mathbf{r}')$ is proportional to the radiant or luminous flux emitted per unit area by the sources located at $\mathbf{r}'$, the term $\mathrm{d}^2\mathbf{r}'$ is the area of an elementary patch of the territory around the point $\mathbf{r}'$, and $K(\mathbf{r}, \mathbf{r}')$ is the light pollution propagation function, also known as "point-spread function" (PSF) accounting for the propagation of the light from $\mathbf{r}$ to $\mathbf{r}'$. The integral is extended to the whole territory whose emissions affect the light pollution indicator $B(\mathbf{r})$, which in some cases may span several hundred km around $\mathbf{r}$. A formal, step-by-step derivation of Eq. (1) from basic radiometric principles can be found in Bará and Lima (2018) and Falchi and Bará (2020). The form of the PSF $K(\mathbf{r}, \mathbf{r}')$ depends on the type of indicator, and on several geometric and spectral quantities. It shall be calculated for the specific state of the atmosphere and for the optical and geographical features of the territory through which the light propagates. Specific examples of PSF can be found, among other, in Cinzano and Falchi (2012), Kocifaj (2007, 2018), Duriscoe et al. (2018), Aubé and Simoneau (2018), and Simoneau et al. (2021).

The way for reducing light pollution, that is, for lessening the value of the appropriate light pollution indicator $B(\mathbf{r})$ in Eq. (1), is twofold. $B(\mathbf{r})$ can be reduced (i) by modifying the way light is emitted or propagates, such that the PSF $K(\mathbf{r},\mathbf{r}')$ has smaller values and/or (ii) by reducing the total amount of light emissions in each patch of the territory, $L(\mathbf{r}')\,\mathrm{d}^2\mathbf{r}'$.

The PSF values can be made smaller -to some extent- by means of engineering interventions, for example by reducing the amount of light emitted towards the upper hemisphere in directions close to the horizontal (reducing ULOR), or by changing the sources' spectra such that the relative weight of some spectral components is attenuated. This reduces the slope of the line A in Fig. 2 (right), bringing it closer to the one corresponding to a perfect lighting system, where it attains its lowest possible value (line P in that figure). It is easy to see that this approach, although able to provide some rather modest gains in the short term, quickly leads to a dead-end alley: no further reductions of the PSF are possible once the lighting systems reach their ideal state (utilization factor equal to 1, appropriate timing, and appropriate spectral and angular emission patterns). The fraction of light reflected in the surfaces we wish to illuminate (see sect 2.1) sets a hard limit to the reductions of the value of the PSF than can physically be achieved. Once this limit is reached (line P in Fig. 2), light pollution continues to increase proportionally to the growth of emissions $L(\mathbf{r}')\,\mathrm{d}^2\mathbf{r}'$ from old and new installations, driven by the new uses of light (ornamental, billboards, etc.) and the progressively higher levels of average street illuminances. An example of the growth of the average illuminance in the streets of Great Britain from 1920 to 2020 is described by Fotios and Gibbons (2018).

This coupling between the PSF $K(\mathbf{r},\mathbf{r}')$ and emissions $L(\mathbf{r}')\,\mathrm{d}^2\mathbf{r}'$ is instrumental to explain the fact that the light pollution levels $B(\mathbf{r})$ show an increasing trend worldwide (Sánchez de Miguel et al. 2021; Kyba et al 2017, 2022) despite the important technical improvements made in modern lighting installations. The reductions in $K(\mathbf{r},\mathbf{r}')$, which are always relatively modest due to physical constraints, are quickly overcome by the unbounded increase of artificial light emissions, $L(\mathbf{r}')\,\mathrm{d}^2\mathbf{r}'$.

## 3. Reducing overall light emissions: the way forward for a sustainable night

Considered from a global perspective, setting a limit to the total light emissions is the way to ensure that the negative effects of light pollution are successfully kept at bay (Luginbuhl et al. 2009; Bará et al 2021). Since nowadays these negative effects have already surpassed the acceptable red-lines in many regions of the world, a verifiable program of light emission reductions becomes necessary. The aim is not just to preserve the present (deteriorated) night, but to restore its quality to acceptable levels. This approach is already enshrined in up-to-date legislation, as the European Union's regulation of the Parliament and of the Council on Nature Restoration (European Union 2024).

This elicits a basic question, namely what is the quality of the night we want to achieve. This is a social choice issue that bears a fundamental importance, not secondary to other possible considerations driven by commercial or industrial interest. The relevance of dark skies for the cultural and scientific heritage of humankind is widely acknowledged, and the key importance of dark nights for biodiversity is nowadays unquestioned. A goal-based policy approach is required to cope with present-time environmental challenges.

By how much should we reduce the artificial light emissions? This is contingent on the answer to the question raised above about the quality of the night, and on the assessment of the present light pollution conditions in the concerned territory. While it is difficult to provide a general answer here, we can describe some possible goals and the reductions that they would require.

Let's consider, for example, the justifiable purpose of retrieving the starry sky in our cities and peri-urban areas. Being able to discern the presence of the Milky Way at the zenith is a desirable goal. What light emission reductions are necessary for that? The answer depends on the present emission levels and on by how much have they deteriorated the night sky in the city of interest. The brightness of the night sky in urban settings, for constant atmospheric conditions and a given lamp technology mix, essentially depends on the density of urban emissions (measured in million of lumen emitted per square km, Mlm/km$^2$) in the first km around the observer (Bará et al. 2021b). Reducing their own city emissions is then a practicable way for municipalities wishing to attain the desired results.

For the purposes of this calculation, we can use the artificial brightness of the zenith sky computed in mcd/m$^2$ by Falchi et al. (2016) in the *New World Atlas* dataset as an indicator of the present conditions at each location of interest. According to the magnitude to luminance conversions used in that work, a pristine natural sky of 22.0 mag$_V$/arcsec$^2$ would have a luminance 0.17 mcd/m$^2$. Spotting the Milky Way at the zenith with the unaided eye requires skies of total brightness (natural plus artificial) ~19.5 mag$_V$/arcsec$^2$. With these conversions this would correspond to a total luminance 1.71 mcd/m$^2$. Subtracting from it the luminance of the natural sky, the maximum allowable artificial zenith luminance turns out to be 1.54 mcd/m$^2$. This target luminance can be compared against the present values, as calculated by Falchi et al. (2016) in pixels 30 arc-second wide spanning the surface of our planet. The ratio between the target and present zenith sky luminance in each pixel is a very close approximation to the ratio between the required and the present whole city emissions, to allow seeing in that pixel the Milky Way with the unaided eye. As a minor note, the widely used conversion between mag$_V$/arcsec$^2$ and luminance proposed by Allen (1973) and applied in Falchi et al. (2016) slightly underestimates the actual luminance of typical night skies, but the suitable correction (Bará et al. 2020) typically modifies the quoted luminances by a CCT-dependent factor of order ~1.17 that does not affect to the values of the indicated ratios.

From these ratios the required amount of emissions reductions can be easily calculated for any city. Figure 3 presents an example of application for A Coruña (43.365°N 8.410°W), a densely built municipality (37.83 km$^2$, 250 000 inhab.) located in the North-West coast of Galicia, an autonomous community nowadays belonging to the kingdom of Spain. The insets show the expected results after different emissions reductions, indicated in the legends as percents (0%, 25%, 50%, 70%, and 75%) to subtract from the initial, present values. The region highlighted in white corresponds to the area where the Milky Way still would not be visible to the unaided eye (the 0% inset describes the present situation, with no reduction). The region highlighted in intermediate gray is the area where the visibility of the Milky Way at the zenith could be successfully restored by means of the indicated reductions. It can be seen that by cancelling the 75% of present emissions the visibility goal could be attained in all the displayed territory. Similar results can be obtained for other cities of this size in many places of Southwest Europe. For attaining full restoration in the metropolitan area of cities of order $10^6$+ inhabitants larger reductions would be required, but a 75% would restore a very significant fraction of their whole territory.

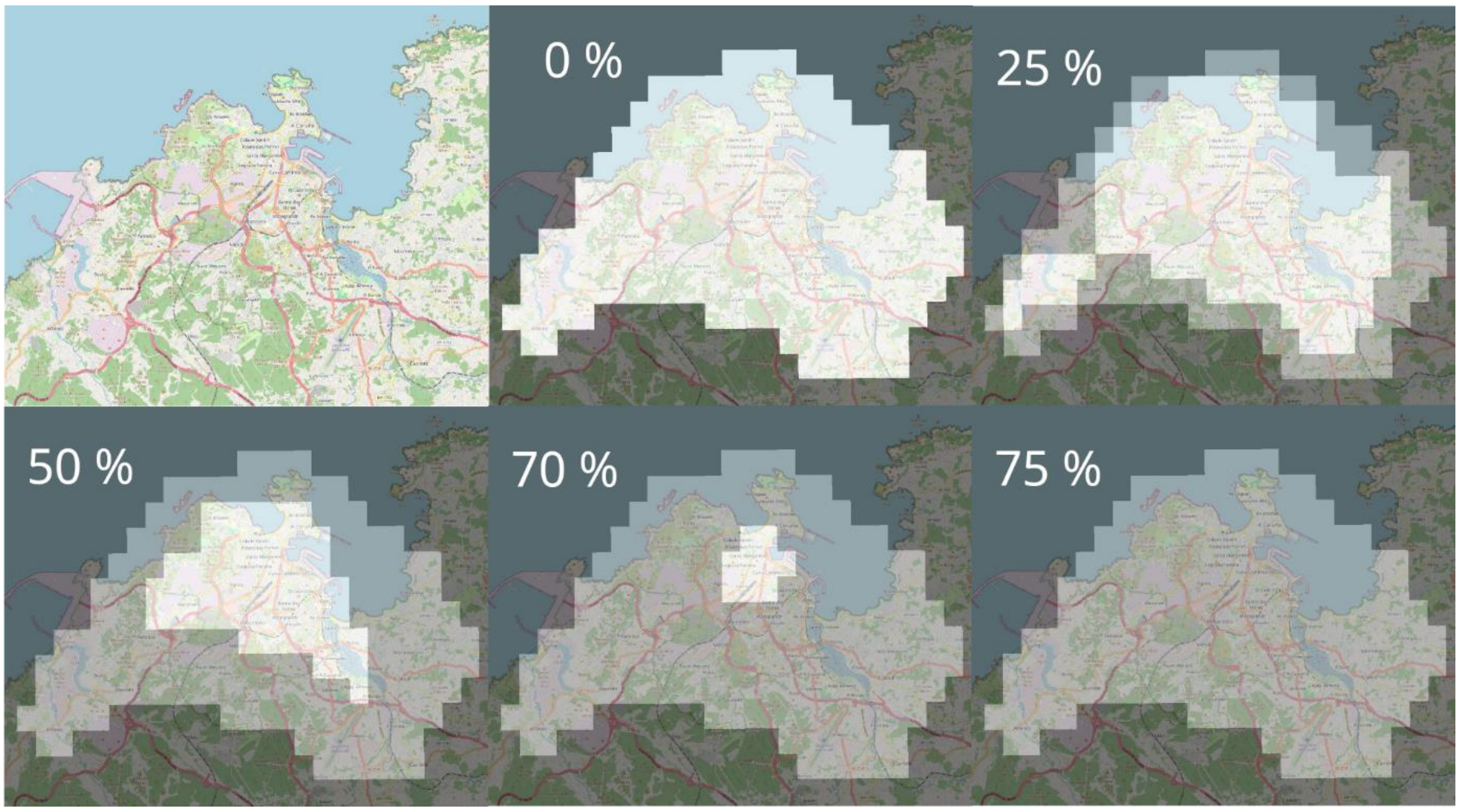

**Fig. 3.** Example of calculation of the light emissions reductions required to restore the ability of seeing the Milky Way at the zenith with the unaided eye. *Upper left*: map of A Coruña (43.365°N 8.410°W) and its surroundings, a densely built municipality of 37.83 $km^2$ and 250,000 inhab. The map spans an area of 17.5x14.5 $km^2$. The remaining insets show the results of different reductions of the total light emissions in the displayed territory. The reductions in the legends are indicated in percents (0%, 25%, 50%, 70%, and 75%) to subtract from the present values. The 0% panel is the initial situation (no reductions). Highlighted in white, area in which the Milky Way would still not be visible. In intermediate gray, area where the Milky Way is not visible today, but would be visible after carrying out the indicated reduction. A 75% emissions reduction would restore its visibility in all the territory. Background map from OpenStreetMap (CC BY-SA 2.0), https://www.openstreetmap.org/. Insets elaborated with QGIS vers. 3.22.10-Białowieża, https://qgis.org/, and GIMP vers. 2.10.36, https://www.gimp.org/.

A similar approach can be adopted to determine the necessary emissions reductions to recover the natural state of the night in isolated astronomical observatories or natural protected areas like those belonging to the Natura 2000 network (https://www.eea.europa.eu/themes/biodiversity/natura-2000/the-natura-2000-protected-areas-network). The main difference in the calculation procedure with respect to the urban case stems from the fact that the light pollution indicators inside each Natura 2000 zone tend to depend significantly on the artificial light emissions produced in a wide region surrounding it, in addition to the emissions produced inside. In cities, the light pollution values are mostly determined by the huge amount of intense sources located within their own borders, with an actual, but generally much smaller contribution from other populated nuclei around. In natural and protected areas, however, the inner emissions tend to be smaller than or comparable to the ones in their surrounding territory, the latter having a non-negligible weight in the light pollution indicator values. The reductions to achieve the restoration goals in these areas should in most cases be extended outside their borders. The extent of the territory where a reduction would be necessary can be easily estimated by applying the methods described in Bará and Lima (2018) and Bará et al (2021). Specific implementations have been published by Aubé et al. (2020), Falchi and Bará (2020, 2021), and Falchi et al (2023b).

## 4. The verification challenge

Accurate and precise monitoring of artificial light emissions is a fundamental task for supervising the outcomes of the public policies aimed to control or reduce light pollution. It is also a serious challenge for light pollution science and technology.

The total emissions of a territory in any given year can be estimated -when available- from public lighting inventories, using site-dependent correction coefficients to account for the fraction of unregistered lights. Inventories of street and ornamental lights are available in many municipalities, sometimes complemented by official registries of other light source types subjected to administrative approval, like commercial lights and advertising billboards. The relative contribution of the remaining, unregistered light sources to the total emissions could in principle be estimated via appropriate sampling on the ground (for an example of a citizen science tool, see Gokus et al (2023)). Trustworthy inventories are of paramount importance for assessing the evolution of the installed lights. Administrative information, however, is not always updated with the frequency and accuracy necessary for enabling the successful inspection and control of the actual emissions of light.

Independent verification is always advisable. It provides a double-check of whether the public policies on lighting are achieving the desired results, and allows to propose corrective measures when needed. There is nowadays a wide toolbox of relatively low-cost instruments and procedures for measuring relevant photometric and radiometric quantities, as described by Hänel et al. (2018) and Barentine (2022). Together with satellite nighttime imagery, they provide invaluable information about our world at night.

We face, however, a problem relative to the accuracy and precision of these remote sensing techniques. Artificial light emissions are growing at rates of order ~2% to 9% per year, depending on the type of light sources and the approach used for the measurements (Kyba et al. 2017, 2023). In order to successfully scrutinize the evolution of the emissions on a yearly basis at an average location of the planet, one should be able to reliably detect emission changes of order ~1% per year. However, the precision of most of the present measurement approaches is worse than that by an order of magnitude. This is due, among other factors, to the unavoidable variability imposed on the measurements by the fluctuations of the state of the atmosphere. The usual methods to monitor the emission of the sources (F) consist on measuring the radiance reaching ground observers (O) or satellites above the terrestrial atmosphere (S) through the direct lines of sight FO and FS (Fig. 4), respectively, or in recording the scattered radiance produced through multiple paths FVO, which is perceived by the observers as artificial sky brightness or skyglow. The fluctuations of the atmospheric extinction and scattering coefficients produce a significant variability in the measurements that cannot always be compensated by averaging large sets of measurements across the year, since there is a non-negligible residual variation of the annual mean atmospheric parameters from one year to the next.

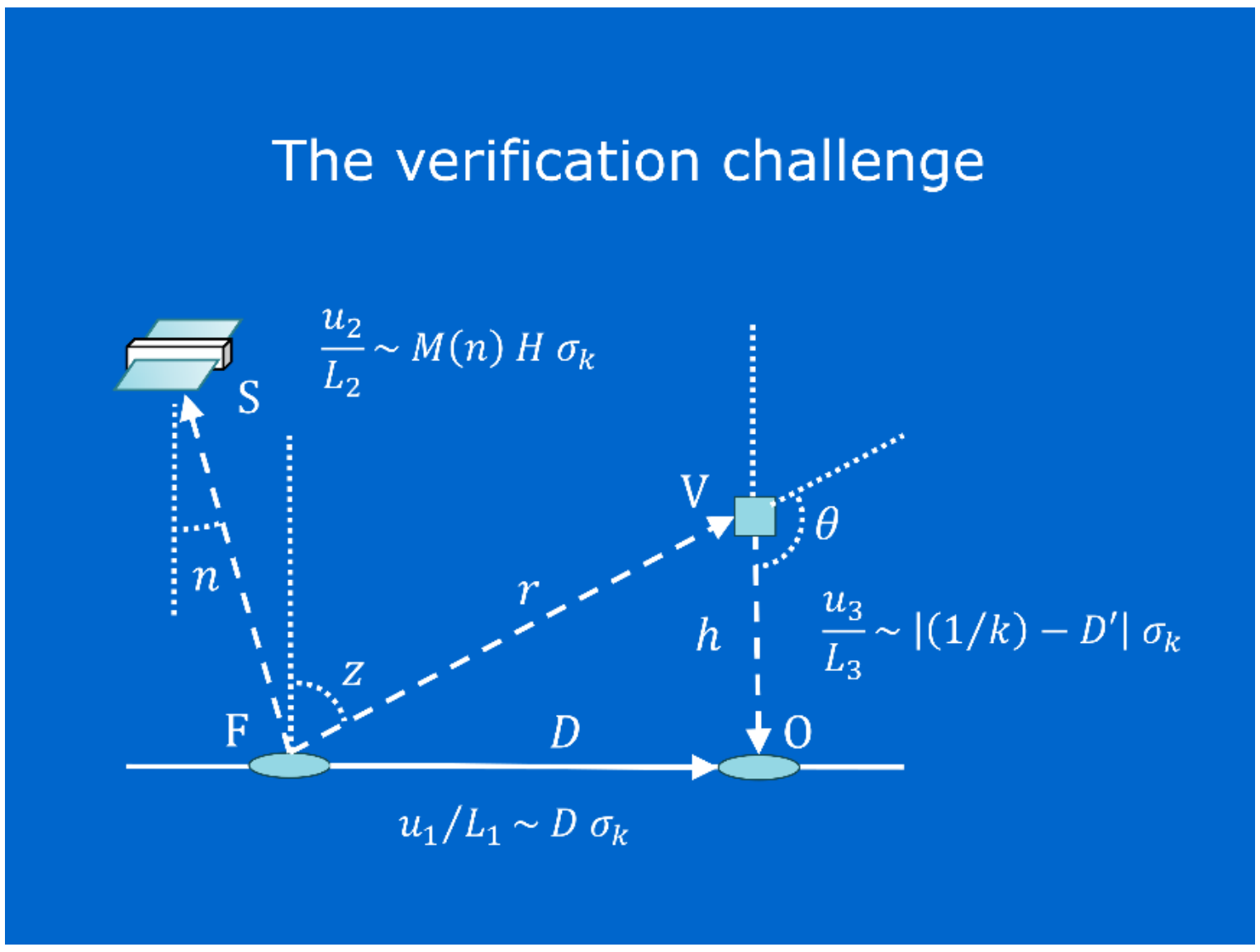

**Fig. 4.** The terrestrial atmosphere is a medium interposed between the light sources (F) and the measuring instruments located on ground (O) or in satellites (S). The fluctuations in the state of the atmosphere set a limit on the minimum relative change of emissions that can be confidently detected.

Whereas large changes in emissions accumulated over several years, of order ~15%, can nowadays be detected with some reliability, detecting changes ten times smaller is a must for monitoring emissions on a yearly or biannual basis, allowing for rapid corrective interventions in case of need. A quantitative study of this and related issues can be found in Bará (2024). One way to improve the precision of the emission estimates is to correct the individual measurements for the actual state of the atmosphere at the time of data acquisition. A recent example of how can this be accomplished is described by Wallner et al (2023). Devising reliable procedures for simultaneous measurement of radiometric and relevant atmospheric variables is one of the next frontiers of this field of research.

## 5. Final remarks

This paper focused on outlining some physical facts that any policy of restoration of the night should be consistent with, without delving into the analysis of the social and political constraints that any light pollution reduction approach is expected to face. It is anyway reasonable to ask ourselves whether emission reductions of the size described in section 3 can be achieved in present industrial societies. The short answer is that this is essentially a matter of choice, whose outcome will be contingent on the political will and collective ability to change the prevalent values about the (ab)use of artificial light outdoors.

There is no fundamental technological difficulty to carry out substantial emission reductions by means of a flexible and gradual transition process. Modern lighting systems, in particular those based on solid-state light sources, may regulate their emissions with ease. How much light do we really need? To put it in historical perspective, the average illuminance in the streets of Britain was ~2 lx at the 1950s, ~5 lx at the late 1960s, ~10 lx in the early 1980s and ~13 lx near the turn of the century (Fotios and Gibbons 2018, Fig. 2). Informal estimates by qualified lighting practitioners suggest that the street illuminance in the cities of the kingdom of Spain could easily average nowadays to ~20 lx. It is well-known that this steady increase in the amount of light in our streets cannot be easily traced back to any reliable, objectively quantified need. Could we live today with the levels of street illumination of the early 1970s? If so, emissions reductions of order ~70%-75% with respect to the present levels do not seem out of reach.

Besides that, large values of street illuminance have been purportedly justified based on traffic safety considerations, specially in conflict areas where pedestrians and drivers share the public space. Their actual efficacy, however, is nowadays under discussion (see, among other, Marchant 2017; Marchant and Norman 2022). New concepts of urban mobility and the reduction of the speed and use of private vehicles pushed today by a growing number of municipalities provide a natural way for adopting P lighting classes (pedestrians / low speed areas) in which values of 5 lx and smaller are not uncommon. Finally, the overall reduction of emissions does not need to be applied uniformly to all city streets; what matters is controlling the total amount of emissions. Higher illuminance levels in some urban districts can be compensated for by larger emission reductions in areas where the light is actually unnecessary.

There are several basic guidelines that shall be adopted in any effective plan for the restoration of the night. Among them, the light emissions reductions should be gradual in time (with implementation periods that may span several years), reasonably homogeneous across the territory, applied to both public and private lights, and respectful of social justice. A detailed discussion of these topics exceeds the limits of this work.

In summary, at the end of day, the restoration of the night will depend more on adopting a consistent strategy for emissions degrowth than on any short-term techno-solutionist approach.

## References

The scientific body of knowledge about light pollution is growing at a fast pace. The references cited in this text are only a small sample of recent works that are directly related to the contents of this article. The interested readers are encouraged to explore the comprehensive "Artificial Light At Night Research Literature Database (ALANDB)" (https://alandb.darksky.org/), curated by J. Barentine, C.C.M. Kyba and A. Wilkerson, an up-to-date catalog of peer reviewed literature about artificial light at night.